\documentclass[12pt]{iopart}
\usepackage[utf8]{inputenc}
\usepackage[english]{babel}
\usepackage{amsfonts}
\usepackage{amssymb}
\usepackage{graphicx}
\usepackage{lmodern}
\usepackage{mathrsfs}
\usepackage{lineno}
\usepackage{hyperref}
\usepackage{color}
\usepackage[dvipsnames]{xcolor}

\newcommand{\nn}{\nonumber}

\def\keywords#1{\vspace{10pt}
     \begin{indented}
     \item[]\rm Keywords: #1\par
     \end{indented}}

\begin{document}



\title{Unruh effect detection through chirality in curved graphene}
\author{J. Madrigal-Melchor$^{1}$, Jairo Mart\'inez--Montoya$^{1,2}$, Alberto Molgado$^{2,3}$ and J. R. Suárez-López$^{1}$}

\address{$^{1}$ Unidad Acad\'emica de F\'isica, Universidad Aut\'onoma de Zacatecas,  \\ 
Calzada Solidaridad esq.~Paseo La Bufa S/N, Zacatecas, Zac, 98060 Mexico}
\address{$^{2}$ Facultad de Ciencias, Universidad Aut\'onoma de San Luis 
Potos\'{\i}, \\
Av.~Salvador Nava S/N Zona Universitaria, San 
Luis Potos\'{i}, SLP, 78290 Mexico}
\address{$^3$ Dual CP Institute of High Energy Physics, Mexico}

\eads{
\mailto{\textcolor{blue}{jmadrigal.melchor@fisica.uaz.edu.mx}},\ 
\mailto{\textcolor{blue}{molgado@fc.uaslp.mx}}\ 
\mailto{\textcolor{blue}{jrsuarez@fisica.uaz.edu.mx}}\
}

\begin{abstract}
We analyze a generalization of the analogue Unruh effect based on curved graphene. To this end, we consider the fourth order in derivatives field theoretic version of the Pais-Uhlenbeck oscillator, for which the Unruh effect may be interpreted as the creation of two different particles with different masses, corresponding to two Klein-Gordon subsystems. For our model, unlike the standard case, electron chirality on the graphene sheet plays a main role as chirality is 
essential to distinguish the couple of particles predicted by the Unruh effect associated to the Pais-Uhlenbeck field model.
\end{abstract}


\section{Introduction}

Some effects predicted by Quantum Field Theory and General Relativity are, until now, simply unreachable in the existing laboratories mainly due to the high  energies required for its observation. In this way, the lack of technology necessary to carry out reliable experiments which could corroborate the phenomena predicted by a theoretic model, has made the scientists to look into alternative ways to verify 
either an emerging effect or a whole theory. Within the gravitational context, the set of alternative proposals are known as Analogue Gravity, which basically consists in the implementation of experimental arrangements of physical systems with properties or symmetries that can reflect a similar behavior with a gravitational theoretical model~\cite{barcelo},~\cite{barcelo2}~\cite{faccio}. There are several attempts to realize  experiments that mimic certain aspects of gravitational phenomena, at both classic and quantum levels, that have been proposed in diverse areas of physics, such as acoustics, optics, and solid state physics~\cite{barcelo},~\cite{barcelo2}~\cite{faccio},~\cite{acoustichawking},~\cite{condensedmattergraphene}. 

Since its synthesis in 2004~\cite{novoselov}, 
graphene has been largely studied due to the wide 
variety of interesting and unusual properties that possesses.  In consequence, graphene has been implemented in many areas of physics in order to test a broad variety of theories and also to explore new phenomena, converting graphene in a potential candidate for a laboratory of the Universe \cite{weylsymmetryIorio}. For this reason, there are several proposals involving the use of graphene as a tool to show some effects in field theories like quantum electrodynamics and gravitation~\cite{chiraltunneling},~\cite{horavagraphene}~\cite{grapheneparallelgravity},~\cite{emergentgravitygraphene},~\cite{gravitationallensgraphene}. Recently, graphene has been proposed as a plausible scenario where Hawking and Unruh-like effects could be observed. Following Iorio, we will focus our attention to the analogue Unruh effect taking place in a sheet of curved graphene~\cite{Iorioreview},~\cite{Ioriofull},~\cite{Iorioshort}.  On the one hand, we will manifestly take advantage of the Weyl symmetry of the action which describes an electron moving on the graphene sheet in order to obtain a comparison with the Unruh effect emerging from the higher derivative model of our interest, that is, the 
field theoretical version of the Pais-Uhlenbeck oscillator. By exploiting Weyl symmetry we can bend the graphene sheet in such a way that it can be described by a metric conformally equivalent to the Rindler metric~\cite{weylsymmetryIorio}. On the other hand, we will imprint a relativistic character to this action  by considering the motion of the electron around the Dirac points associated to the electronic structure of graphene. As described below (also see~\cite{Iorioreview},~\cite{Ioriofull},~\cite{Iorioshort}), this considerations set the scenario where the analogous Unruh effect will appear. 

In this way, the main purpose of this article is to introduce the analogue  effect appearing in a curved graphene layer in comparison to the Unruh effect for a higher order derivative field theory associated to the Pais-Uhlenbeck oscillator. The field theoretical Pais-Uhlenbeck model serves as a toy model within the context of conformal gravity, an alternative theory  describing the gravitational field that naturally incorporates the Weyl symmetry.    The Unruh effect for the Pais-Uhlenbeck field  model was recently  described in~\cite{PUunruh} and, as discussed there, contrary to the standard Unruh effect described from the massless Klein-Gordon model, the Pais-Uhlenbeck field depends explicitly on the masses (and thus in the  frequencies) defining the model, hence a different treatment must be implemented in order to properly describe the Unruh effect. In particular, the Unruh effect for the Pais-Uhlenbeck model emerges in the unequal frequencies limit and the evidence presented in~\cite{PUunruh} suggests that the effect is absent in the equal frequencies limit. Bearing this in mind, we will focus our attention  to the different frequencies case, and we will include comments on the equal frequencies limit where appropriate.  Within this case, as reviewed below,  the Unruh effect  predicts the emergence of couples of particles which may be interpreted as particles and antiparticles of different masses. This interpretation is physically consequent with the equal frequencies limit, since for this case the emerging particles and antiparticles acquire equal masses  thus annihilating each other.  In this regard, our proposal rests on the study  of the analogue Unruh effect on a curved graphene sheet via the chirality of  electrons lying  in different Dirac points and, in agreement with reference~\cite{masschiralitybreak}, we will introduce chirality of  graphene  as a mechanism to break the symmetry which is equivalent to incorporate different masses in electrons  needed in order to distinguish the 
analogous couple of different particles appearing in the Unruh effect for the Pais-Uhlenbeck field model.

The rest of the article is organized as follows. In Section \ref{sec:UnruheffectPU} we start by concisely describing the thermal behavior of particles associated to the Unruh effect for the Pais-Uhlenbeck field model. In Section \ref{sec:analoguePU} we give a brief summary of the analogue model on a graphene layer of the Unruh effect for the Pais-Uhlenbeck  model and discuss the new properties that emerge due to the symmetries of this model and the relevance that these properties play in order to describe the Unruh effect. Finally, in Section \ref{sec:conclusions} we include some concluding remarks.  

\section{Unruh effect for the Pais-Uhlenbeck  field model}
\label{sec:UnruheffectPU}

In this section we will  introduce the field theoretic version of the Pais-Uhlenbeck oscillator and briefly review the thermal behavior associated to the Unruh effect for this model.
To this end, we will follow as close as possible reference~\cite{PUunruh}.
 The Pais-Uhlenbeck model is described in a four dimensional Minkowski spacetime with signature $(+---)$ and local coordinates denoted by $x$. 
The action for this model is
\begin{eqnarray}\label{action pu}
S = \int \phi(\Box + m_1^2)(\Box + m_2^2)\phi \, d^4x \,,
\end{eqnarray}
where $\phi=\phi(x)$ denotes a scalar field, $m_1$ and $m_2$ are free parameters usually identified with the masses of the Pais-Uhlenbeck field model, and $\Box=\partial_{\mu}\partial^{\mu}$ is the d'Alembert operator. The field equation governing this model 
simply reads
\begin{eqnarray}\label{pu eq}
(\Box+m_1^2)(\Box+m_2^2)\phi = 0 \,.
\end{eqnarray}
By symmetry, one may argue that the solution to equation (\ref{pu eq}) is the linear combination~\cite{zwillinger}
\begin{eqnarray}
\label{eq:PUsols}
\phi(x) = \alpha u(x) + \beta v(x) \,,
\end{eqnarray}
where $\alpha$ and $\beta$ are arbitrary constants, while $u(x)$ and $v(x)$ are complex linearly independent solutions to the Klein-Gordon equations with masses $m_1$ and $m_2$, respectively. For solutions given by the equation~(\ref{eq:PUsols}), we may define a genuine inner product by means of the Noether current associated to the action~(\ref{action pu})
\begin{eqnarray}
\label{eq:innerp}
(\phi,\psi) &:=& \int_\Sigma i\left[ -(m_1^2+m_2^2)(\phi^*\partial_{\mu}\psi-\psi\partial_{\mu}\phi^*)+(\partial_{\mu}\phi^*\Box\psi-\partial_{\mu}\psi\Box\phi^*) \right. \nn\\
& &\left. -(\phi^*\partial_{\mu}\Box\psi-\psi\partial_{\mu}\Box\phi^*)\right] d\Sigma^{\mu} \,,
\end{eqnarray}
where $d\Sigma^{\mu}:=n^{\mu}d\Sigma$, being $\Sigma$ 
a spatial Cauchy hypersurface and $n^{\mu}$ a unitary vector orthogonal to $\Sigma$.  Also, $d\Sigma$ stands for the volume element of the hypersurface $\Sigma$. 
This inner product is obtained from the Pais-Uhlenbeck 
Lagrangian and, thus, it contains the symmetric properties of the system.  Also note that the construction of 
this inner product results in complete agreement with the  
definition of the inner products for standard fields in Quantum Field 
Theory~\cite{ryder}. 
One may easily verify that under this inner product the independent set of solutions $\{u(x)\}$ and $\{v(x)\}$ result orthogonal.  
We now perform a formal expansion of the field in terms of the families of complex classical solutions, $\{u(x)\}$ and $\{v(x)\}$, and the annihilation and creation operators, in the following way
\begin{eqnarray}
\phi(x)=\sum_{i=0}^{\infty}\left[u_i(x)a_i + u_i^*(x)a_i^{\dagger} + v_i(x)b_i + v_i^*(x)b_i^{\dagger}\right]\,,
\end{eqnarray} 
where $a_i$ and $b_i$ are the annihilation operators and $a_i^{\dagger}$ and $b_i^{\dagger}$ are the creation operators, satisfying the commutation relations
\begin{eqnarray}
\label{eq:basiccomm}
[ a_{\vec{k}},a_{\vec{k}'}^{\dagger} ] 
= 
\delta(\vec{k}-\vec{k}')
=
- [ b_{\vec{k}},b_{\vec{k}'}^\dagger ] \,,
\end{eqnarray}
and otherwise vanishing. Now we look for a Poincaré invariant extension of the standard two-point function, $W(x,x')=\left<\Omega\right|\phi(x)\phi(x')\left|\Omega\right>$.  In the case of our interest, this generalization of the two-point function
is necessary, according to the Stone-von Neumann theorem, in order to guarantee the maximum number of representations for our model~\cite{moschella1},~\cite{moschella2}. Due to the symmetry exhibited by  the solution of the field equation, and the fact that the solutions $u(x)$ and $v(x)$ are orthogonal within the inner product introduced above in~(\ref{eq:innerp}), the two-point function can be decomposed as
\begin{eqnarray}
\label{eq:twopointW}
W(x,x')=W_u(x,x')-W_v(x,x')  \,.
\end{eqnarray}
where each $W_u(x,x')$ and $W_v(x,x')$ represent the two-point function for the modes $u(x)$ and $v(x)$, respectively. To get the exact expressions for the modes we need to solve the Klein-Gordon equations that follow each of the modes. As we are interested in the analysis of the Unruh effect, we will solve the corresponding
 Klein-Gordon equations in Rindler coordinates since this set of coordinates allow us in a natural way to define uniformly accelerated  observers.  Also, these coordinates will grant us a straightforward comparison with the metric of the curved graphene layer of the next section. Thus, we will consider Rindler coordinates as given by the mapping $(t,x,y,z)\mapsto(\rho\sinh(\eta),\rho\cosh(\eta),y,z)$, and we will focus our attention in the $\eta$ and $\rho$ coordinate sector.  Within this sector, the Klein-Gordon equations read  
\begin{eqnarray}
\left( \frac{1}{\rho^2}\frac{\partial^2}{\partial \eta^2} - \frac{1}{\rho}\frac{\partial}{\partial \rho}\left(\rho\frac{\partial}{\partial \rho}\right) + m_1^2 \right) U_{\omega_n}
& = &
0 \,, 
\qquad
n
=
1,2\,,
\end{eqnarray}
where, for the sake of simplicity,  we have introduced the compact notation $U_{\omega_n}:=(u(\eta,\rho), v(\eta,\rho))$, in such a way that the solutions for each of the $u(\eta,\rho)$ and $v(\eta,\rho)$ modes 
may be written as
\begin{eqnarray}\label{eq:uv-prod-omegas}
U_{\omega_1}=
e^{-i\omega_1\eta}K_{i\omega_1}(m_1\rho)  \,,
\qquad
U_{\omega_2}=
e^{i\omega_2\eta}K_{i\omega_2}(m_2\rho) \,,
\end{eqnarray}
respectively. Here $\omega_n$ stand for the frequencies
of the Pais-Uhlenbeck oscillator~\cite{smilga} which, for a given mode $k$, are related to their corresponding masses $m_n$ via the dispersion relations
\begin{eqnarray}
\label{eq:disp}
\omega_n^2 
= 
k^2 + m_n^2 \,,
\qquad
n
=
1,2\,.
\end{eqnarray}
Also, for any fixed value of the index $n$, the $K_{i\omega_n}(m_n\rho)$ functions appearing in~(\ref{eq:uv-prod-omegas}) represent the Macdonald functions, which are solution to the modified Bessel equation of second kind with imaginary 
index~\cite{atlas}
\begin{eqnarray}\label{eq:mac}
\hspace{-10ex}
\rho^2\frac{d^2 K_{i\omega_n}(m_n\rho)}{d \rho^2} + \rho\frac{d K_{i\omega_n}(m_n\rho)}{d \rho} - \left(m_n^2\rho^2-\omega_n^2\right)K_{i\omega_n}(m_n\rho)=0\,.
\end{eqnarray}
From now on, and without losing generality, we will consider the case $\omega_1>\omega_2$.  
The equal frequencies case, as discussed in~\cite{PUunruh},  must be treated differently as the 
Pais-Uhlenbeck model is governed, contrary to our case, by a 
quantum Hamiltonian with continuous spectrum.  However, by considering a proper limit within our case, one may argue that the Unruh effect will be absent as the probability amplitude vanishes. The difference between the sign in the exponential of the solutions is necessary to guarantee the positiveness of the inner product defined in (\ref{eq:innerp}). 
This difference in signs may be traced back to the non-Hermiticity property of the Pais-Uhlenbeck oscillator. Indeed, an appropriate inner product for non-Hermitian models can be constructed within the $\mathcal{PT}$-symmetric quantum mechanical formalism, 
for example, for which positive eigenvalues
may be assigned to a non-Hermitian Hamiltonian~\cite{bender}. 
By considering the inner product~(\ref{eq:innerp})
one may verify that the normalized modes 
are given by 
\begin{eqnarray}\label{eq:modos uv}
U_{\omega_n}
=
\frac{\sqrt{\sinh(\pi\omega_n)}}{\Delta\pi}e^{-\sigma_ni\omega_n\eta}K_{i\omega_n}(m_n\rho) \,,
\qquad
n
=
1,2 \,,
\end{eqnarray}
where we have introduced the notation $\sigma_1=1$ for the $u$-modes and $\sigma_2=-1$ for the $v$-modes, in agreement with~(\ref{eq:uv-prod-omegas}), and we also defined $\Delta:=\sqrt{m_1^2-m_2^2}=\sqrt{\omega_1^2-\omega_2^2}$, where the last 
equality is obtained by considering the dispersion 
relations~(\ref{eq:disp}). 
As discussed in~\cite{PUunruh}, particles associated to each of the modes (\ref{eq:modos uv}) may be interpreted as particles of mass $m_1$ and antiparticles of mass $m_2$, respectively.  This different interpretation 
for the particles associated to the modes lies on the
value of the $\sigma_n$ appearing in the exponential function.
Further, we note that the equal frequency limit, $\Delta\to 0$, clearly results ill-defined.  Indeed, even at the classical level the Hamiltonian for the Pais-Uhlenbeck oscillator 
results different in nature whenever we consider different or equal frequencies from the beginning.
As discussed in~\cite{bolonek} 
(see also~\cite{smilga} and~\cite{pudeformation}) this 
inequivalence at the Hamiltonian level reflects in different spectrum for the  quantum Hamiltonian operators corresponding to either the different or equal frequencies cases.  Thus, as mentioned before, we will only consider here the different 
frequencies  $(\omega_1>\omega_2)$ case, and from this perspective  
we will only provide some arguments pointing towards the absence of the Unruh effect in the equal frequencies limit. Note then that, within this perspective, our physical interpretation as assigning a couple particle and 
antiparticle for the different modes, results 
consequent with the equal frequencies limit as, within this limit, the particles and antiparticles annihilate each other, causing no observation of particles created.
See~\cite{PUunruh} for further discussion and 
details.

For each of the normalized modes $U_{\omega_n}$~(\ref{eq:modos uv}), 
the generalized Poincaré invariant two-point function $W_{U_{\omega_n}}(x,x')=\left<\Omega\right|\phi(x)\phi(x')\left|\Omega\right>$  thus reads 
\begin{eqnarray}\label{funcion modos uv}
\hspace{-14ex}
W_{U_{\omega_n}}(x,x') = \frac{\sigma_n}{\pi^2\Delta^2}\int_0^{\infty} \sinh(\pi\omega_n)\left[ \frac{e^{-i\omega_n(\eta-\eta')}}{1-e^{-2\pi\omega_n}} -\frac{e^{i\omega_n(\eta-\eta')}}{1-e^{2\pi\omega_n}} \right] K_{i\omega_n}(m_n\rho)K_{i\omega_n}(m_n\rho')d\omega_n \,, \nn \\
\end{eqnarray}
which, after a straightforward algebraic manipulation, may 
be expressed as an inverse Fourier transformation, 
that is,
\begin{eqnarray}\label{eq:general two point functions}
\hspace{-5ex}
W_{U_{\omega_n}}(x,x') = \frac{\sigma_n}{\pi^2\Delta^2}\int_{\infty}^{\infty}e^{i\omega_n(\eta-\eta')}\frac{K_{i\omega_n}(m_n\rho)K_{i\omega_n}(m_n\rho')\sinh(\pi\omega_n)}{e^{2\pi\omega_n}-1} d\omega_n \,,
\end{eqnarray}
for $n=1,2$. 
Thus,  the Fourier transform for each of the two-point functions has the form
\begin{eqnarray}\label{eq:inverseFourier}
F(\omega_n)=\frac{2\sigma_n}{\pi\Delta^2}\, \frac{K_{i\omega_n}(m_n\rho)K_{i\omega_n}(m_n\rho')\sinh(\pi\omega_n)}{e^{2\pi\omega_n}-1} \,,
\end{eqnarray}
where, in a natural way, we may identify the Planck factor, $(e^{2\pi\omega_n}-1)^{-1}$ for each of the emerging particles, also telling us that any of the created particles follow a Bose-Einstein distribution, and the temperature is related  to their respective frequencies $\omega_n$, as discussed in~\cite{moschella1},~\cite{moschella2}.  The function
$F(\omega_n)$ 
is precisely the one that will serve to 
establish the analogy with the graphene layer as 
in that case the power spectrum, that is, 
the Fourier transform of the vacuum expectation value,  will represent the 
particles as seen from an accelerated observer.
Hence, in the following section we will review the analogue Unruh effect in graphene and 
we will also implement this analogy for the model of our
interest.

\section{Analogue Unruh effect for the Pais-Uhlenbeck field model}
\label{sec:analoguePU} 

In order to describe the context in which the Unruh effect takes place in graphene we will follow~\cite{Iorioreview},~\cite{Ioriofull}. The starting point will be to consider graphene as 
a relativistic spacetime object.  This, as reported in the literature,  will allow us to find analogies with either relativistic quantum field theory, gravitation or 
high energy physics.   Indeed, 
as it is well known, graphene may be thought of as a relativistic entity due to the fact that electrons moving in a vicinity of the so called Dirac 
points are described by a linear dispersion relation, 
thus behaving  as massless Dirac particles with a chiral nature \cite{katsnelson2012}.
Besides, 
the metric on a curved graphene layer (that is, a function that allow us to introduce a notion of distance among points in the layer) may be expressed as a 
spacetime $(2+1)$-dimensional diagonal metric by considering 
\begin{eqnarray}\label{eq:graphene metric}
g_{\mu\nu}^{\mathrm{graphene}} = \left( \begin{array}{cc}
1 & 0 \\
0 & g_{ij}
\end{array} \right) \,,
\end{eqnarray} 
where the first entry corresponds to the temporal component of the metric, and $g_{ij}$ stands for the $2$-dimensional standard metric on the graphene sheet. In this sense, in order to reproduce a curved spacetime,
we can deform the graphene sheet, thus bringing into play the curvature of the metric.  In our case the curvature will be present only in the spatial part. Altogether, since electrons moving on the graphene are described by Dirac equation, Weyl symmetry can be used in order to map the graphene metric 
into a Rindler spacetime by means of a conformal 
transformation~\cite{weylsymmetryIorio}. This conformal symmetry will 
be essential to establish the analogy of our interest 
with the Pais-Uhlenbeck field model, which, as discussed before, serves as a toy model for conformal Weyl gravity.

In order to show that the graphene 
metric~(\ref{eq:graphene metric}) is conformal to 
Rindler spacetime, we will start by considering 
the graphene sheet as a surface of revolution described by the line element
\begin{eqnarray}
dl^2 &=& g_{ij}dx^idx^j=da^2 +R^2(a)db^2 \, ,
\end{eqnarray}
where $a$ and $b$ are the meridian and parallel coordinates, respectively, while $R(a)$ is the curve from which the surface of revolution is generated. For the specific case of our interest, the curve is chosen to be $R(a)=c\exp(a/r_0)$ in order to 
reproduce a Beltrami pseudo-sphere, where $c$ is an arbitrary constant, and  
$r_0$ is a positive 
parameter associated to the curvature radius and related to the Gaussian curvature of the surface by means of 
$\mathcal{K}=-(1/r^2_0)$. By introducing harmonic coordinates\footnote{Recall that harmonic coordinates 
are defined as those which are solutions to Laplace-Beltrami equations and, as it may be seen in any standard reference, they may be thought of as an approximation to the concept of inertial frame of reference from the general relativistic point of view~\cite{waldrelativity}.}, $(\tilde{x}(a,b),\tilde{y}(a,b))$, explicitly defined as
$\tilde{x} = b/r_0$ and $\tilde{y}= (1/c)\exp{(-a/r_0)}$,
the metric on the graphene surface reads
\begin{eqnarray}
\label{eq:conformalmetric}
dl^2 &=& \frac{r_0^2}{{\tilde{y}}^2}\left( d{\tilde{x}}^2 + d{\tilde{y}}^2 \right)\,,
\end{eqnarray} 
and such that the coordinate  $\tilde{y}$ 
is restricted to the plane $\tilde{y}>0$. 
At this point it is worth mentioning that the introduction of harmonic coordinates is relevant from our perspective for two reasons~\cite{smith}:  on the one hand, 
one may show that for any surface of revolution these coordinates can always be explicitly found
and, on the other hand, the metric generated by these coordinates results conformally equivalent to the 2-dimensional Euclidean metric, as seen in~(\ref{eq:conformalmetric}). Thus, from~(\ref{eq:graphene metric}), the complete spacetime metric for the graphene may be written as
\begin{eqnarray}\label{eq:harmonic metric}
ds^2 &=& g_{\mu\nu}^{\mathrm{graphene}}dx^\mu dx^\nu= dt^2-dl^2  = \frac{r_0^2}{{\tilde{y}}^2}\left( \frac{{\tilde{y}}^2}{r_0^2}dt^2 - d{\tilde{x}}^2 - d{\tilde{y}}^2 \right)\,,
\end{eqnarray}
or, in terms of the original meridian and parallel coordinates, $a$ and $b$, respectively, the metric (\ref{eq:harmonic metric}) acquires the form
\begin{eqnarray}
ds^2 &=& \frac{c^2}{r_0^2}e^{2a/r_0}\left[ \frac{r_0^2}{c^2}e^{-2a/r_0}\left( dt^2 - da^2 \right) - r_0^2db^2 \right]\,,
\end{eqnarray}
where the expression inside the square bracket may be
recognized as the $(2+1)$-dimensional Rindler metric $ds_{\mathrm{Rindler}}^2= Y^2 dT^2-dX^2-dY^2$ by
performing the trivial identification $T=t/r_0$, 
$X=r_0b=r_0^2\tilde{x}$  and  $Y=\pm(r_0^2/c)\exp(-a/r_0)=\pm r_0^2 \tilde{y}$~\cite{waldrelativity}.  In this way we conclude that 
the graphene metric is related to the Rindler spacetime via a conformal transformation, with conformal factor 
explicitly given by $[(c/r_0)\exp(a/r_0)]^2$. This is the first analogy between the graphene and the 
Pais-Uhlenbeck studied in the last section. The second analogy between these two systems emerges
by considering the power spectrum (also referred to as 
the response function in the Quantum Field Theory lore) for both,  
the 
field theoretical Pais-Uhlenbeck model and the 
graphene layer. On the one hand, the power spectrum represents the particles that an accelerated observer will experience~\cite{birrell} and,
in our notation, it is represented by the function
\begin{eqnarray}\label{eq:power spectrum}
F(\omega_n) = \int_{-\infty}^{\infty} e^{-i\omega_n\eta}W_{U_{\omega_n}}(x,x')d\eta \,.
\end{eqnarray}
Here $\eta$ stands for the temporal component of 
Rindler coordinates and $W_{U_{\omega_n}}(x,x')$ being the generalized 
two-point function obtained in~(\ref{eq:general two point functions}) and, as discussed there,  associated to the expectation value of the vacuum.  In this way, 
the power spectrum $F(\omega_n)$ for the Pais-Uhlenbeck model is simply given by relation~(\ref{eq:inverseFourier}).  In comparison, 
we may note that in 
the standard case of a scalar massless Klein-Gordon field the power 
spectrum is obtained by considering the positive frequencies of the Wightman function instead of our generalized 
two-point function $W_{U_{\omega_n}}(x,x')$ (see~\cite{birrell} and~\cite{takagi} for further details in this standard case). 
On the other hand, for a graphene sheet the local density of states represents the number of accessible states  per unit of energy on a lattice site, in particular at two different Dirac points.  As described 
in~\cite{cortijo},~\cite{altland}, for the case of graphene  the local density of 
states $\rho(\omega,x)$ is linearly proportional to the power spectrum. Thus
\begin{eqnarray}\label{ldos}
\rho(\omega,x) \approx F(\omega,x)\,,
\end{eqnarray} 
where $x$ corresponds to any of the two 
different Dirac points. 
In this manner, as we were able to deform the graphene layer in order to manifestly show the Rindler
geometry and, also, as the local density of states
is directly related to the power spectrum, following~\cite{weylsymmetryIorio},~\cite{Iorioreview},~\cite{Ioriofull},~\cite{Iorioshort}, thus we establish 
the setting on which the analogy between the 
graphene and the Unruh effect for the Pais-Uhlenbeck field will rest.

In order to start our analogy we recall that,
contrary to the standard massless scalar Klein-Gordon case,  
for the field theoretical Pais-Uhlenbeck model we analyzed the 
thermal behavior of the particles predicted by the Unruh effect by introducing  a generalized two-point function, that is, the most general vacuum expectation value
which is invariant under Poincaré transformations, as described in   
section~\ref{sec:UnruheffectPU} above. This generalized 
two-point function thus corresponds to the Wightman function used in the standard Unruh effect for a massless field.  However, we must emphasize that 
in our case the inherent symmetry of the field of our interest plays a fundamental role as the generalized two-point function was interpreted before as 
corresponding to the creation of two different particles.  More specifically, our model corresponded to two Klein-Gordon subsystems with different non-vanishing masses. 
Therefore, in our case we will propose to use the mode symmetric two-point function~(\ref{eq:twopointW}), and the explicit expressions (\ref{eq:general two point functions}) for any of the $U_{\omega_n}$ modes,
in the power spectrum function~(\ref{eq:power spectrum}), 
and finally, we will introduce both power spectrum 
functions $F(\omega_n)$ in the local density of 
states~(\ref{ldos}) in order to set up the 
formal analogy between both, the Pais-Uhlenbeck model and 
the graphene layer.   In this manner, we find that the 
local densities of states adapted to the system of our 
interest is given by 
\begin{eqnarray}
\label{eq:rho}
\rho(\omega,x) &\approx & F(\omega_1) + F(\omega_2)  
\nn \\
&=& 
\frac{2}{\pi\Delta^2}\,
\sum_{n=1,2} \sigma_n\, \left[\frac{K_{i\omega_n}(m_n\rho)K_{i\omega_n}(m_n\rho')\sinh(\pi\omega_n)}{e^{2\pi\omega_n}-1} \right]
\end{eqnarray}
Note that the minus sign appearing in front of the second term is inherited from our physical interpretation  
of the Pais-Uhlenbeck model, for which particles
of mass $m_1$ and frequency $\omega_1$ and antiparticles of mass $m_2$ and frequency $\omega_2$ are associated 
to the two-point functions $W_u(x,x')$  
and $W_v(x,x')$, respectively.  In consequence, in two different Dirac points within the graphene layer 
this should be interpreted as the detection of two different kinds of electrons
on the graphene sheet, each distinguished by its respective mass. Even though this may sound 
counterintuitive at first sight, it may be understood by taking into account chirality effects on any of these electrons.
Certainly, as demonstrated in~\cite{chiraltunneling}, nonequivalence  of the Dirac points on the graphene lattice 
induces different 
 chirality on electrons, which has been observed in photoemission experiments \cite{Liu2011a}.    
Hence, the topological structure of the graphene layer as composed by two nonequivalent triangular sublattices, guarantees a
reliable distinction among electrons belonging to  different Dirac points.
This feature appearing in 
graphene plays a fundamental role in our proposal, as this allows us  to settle an analogy,  in a legitimate manner, with 
the Unruh effect associated to the field theoretic version of the Pais-Uhlenbeck oscillator.  Indeed, 
as discussed before, the pair of particles emerging 
for the Pais-Uhlenbeck model may be interpreted 
within the graphene as electrons belonging to different triangular sublattices, and distinguished by different 
chirality.  Bearing this in mind, and in agreement 
with~\cite{masschiralitybreak}, we may think of chirality as a mechanism to generate mass for electrons 
in the graphene layer \cite{Semenoff2012}.   
In this way, by incorporating chirality within our 
graphene setup, we are able to establish the analogy 
with the creation of the pair particle and antiparticle 
of different masses for the Pais-Uhlenbeck field theoretical model.  
Further note that experimentally, chirality may be detected by 
measuring the Berry phase of electrons, which in turn 
changes the transport properties of the 
graphene~\cite{kim}. 
This measurement could serve to corroborate our proposal
by emphasizing the relevance of chirality within our model.
Once we have introduced electrons with different 
masses on the graphene layer, we 
may incorporate the original experimental 
setup~\cite{Iorioexperiment}
in order to fulfill the analogy with the 
Unruh effect for the model of our interest.

We also want to emphasize that, as stated by 
Iorio~\cite{Ioriofull} for the massless scalar field, knowledge of the power spectrum might be very helpful in 
order to elucidate the physical content of the vacuum
condensate.  In our case, however, the model we have considered is inherently mass dependent.  This may seem 
problematic as for the massive Klein-Gordon field a
Wightman function results non-analytic (see~\cite{takagi} for further details) and, thus, in order to 
extract information about the 
vacuum we are forced to use asymptotic expansions.  
We circumvent this situation by appropriately 
replacing the Wightman function by the generalized Poincar\'e invariant two-point function described 
in~\cite{moschella1},~\cite{moschella2}.  In consequence, it is from 
this generalized two-point function that we 
obtain the physical information of the vacuum state.

Finally, we may note that the equal frequencies limit
may be absent in the experimental setup mentioned above.  
This issue may be seen from two perspectives.  In the 
first place, relation~(\ref{eq:rho}) results ill-defined
within this limit.  As mentioned in last section, this limiting case should be obtained from a totally different construction.  Secondly, the equal frequency limit simply does not appear relevant, as each cell in  the graphene consists of two interpenetrating triangular 
sublattices and, for any of these sublattices we 
associate electrons with different masses.  
Hence, within the equal frequency case the 
sublattices would be indistinguishable. This situation,
as discussed above, is not possible by considering the 
chirality property for each of the Dirac points.

\section{Conclusions}
\label{sec:conclusions}

In this paper we presented a correspondence of the Unruh effect associated to the field theoretic version of the Pais-Uhlenbeck model and the standard analogue Unruh effect based on graphene.  In this sense, we established 
how the pair of particles of different masses naturally emerging via the Unruh effect within the Pais-Uhlenbeck 
model may be interpreted by incorporating 
chirality effects in two different Dirac points on the graphene layer.   In our proposal, thus, chirality of the electrons resulted very relevant as this physical property allowed us to distinguish electrons in the graphene layer by conferring the electrons an effective mass. 
Chirality emerges in the graphene in a natural way since the Dirac points that conform the lattice are topologically different and, in consequence, electrons belonging to different triangular sublattices acquire different chirality. 
The situation here must be confronted with the analogue Unruh effect emerging in the standard case, that is, for 
the massless Klein-Gordon field, for which electrons are 
thought of as massless particles, thus avoiding the 
introduction of chirality effects to generate an 
effective mass.  
Therefore, as compared with the massless scalar field, the Pais-Uhlenbeck field model introduced 
resulted in a more appropriate scenario to implement the 
Unruh effect in graphene as the natural structure of
graphene consists in two different interpenetrating 
sublattices, each one of them naturally assigned by chirality to the two 
different particles emerging via the Unruh effect for 
the field theoretical model we considered.

It is important to emphasize that implementation 
of the analogue 
Unruh effect proposed here for a graphene layer  
might be generalized, to some extent, within the context of conformal gravity for which the Pais-Uhlenbeck model 
is commonly used as a prototype.   From this point of view, direct verification of the emergence of the Unruh 
effect in the graphene layer as mirroring the 
effect in the Pais-Uhlenbeck model will be very relevant
and it may help to  reconsider this alternate theory 
for the gravitational field.  This will be done elsewhere.

\section*{Acknowledgments}

The authors would like to thank Jasel Berra-Montiel for encouraging discussions and collaboration. AM acknowledges financial support from CONACYT-Mexico under project CB-2014-243433.

\end{document}